# XML CONTENT WAREHOUSING: IMPROVING SOCIOLOGICAL STUDIES OF MAILING LISTS AND WEB DATA

by

Benjamin Nguyen
(PRiSM, Université Versailles St-Quentin et INRIA, Projet SMIS)
Antoine Vion
(LEST, Université de la Méditerranée et LabexMed)
François-Xavier Dudouet
(IRISSO, Université Paris-Dauphine)
Dario Colazzo
(LRI, Université de Paris Sud et INRIA, Equipe LEO)
Ioana Manolescu
(INRIA, Equipe LEO, et LRI, Université de Paris Sud)
Pierre Senellart
(Telecom Paris Tech, Equipe DB Web)

**Résumé :**
Dans cet article, nous présentons les lignes directrices d'une approche basée sur XML pour l'étude sociologique des données Web tels que l'analyse des listes de diffusion ou bases de données disponibles en ligne. L'utilisation d'un entrepôt XML est une solution flexible pour le stockage et le traitement de ce type de données. Nous proposons une solution déjà mise en place et montrons des applications possibles avec notre étude de profils d'experts impliqués dans des actions normatives W3C. Nous illustrons l'utilisation de bases de données sociologiques semi-structurées en présentant notre schéma XML pour le stockage de listes de diffusion. Un schéma XML permet de nombreuses adjonctions ou croisements de sources de données, sans modifier les données déjà stockées, tout en permettant de possibles évolutions structurelles. Nous montrons également que l'existence de données cachées implique une complexité accrue pour les utilisateurs SQL traditionnels. Le stockage par XML permet l'entreposage totalement exhaustif et de requêtes récursives dans le contenu, avec beaucoup moins de dépendance au stockage initial. Nous présentons enfin la possibilité d'exporter les données stockées vers des logiciels avancés couramment utilisés et consacrés à l'analyse sociologique.

**Abstract:**

In this paper, we present the guidelines for an XML-based approach for the sociological study of Web data such as the analysis of mailing lists or databases available online. The use of an XML warehouse is a flexible solution for storing and processing this kind of data. We propose an implemented solution and show possible applications with our case study of profiles of experts involved in W3C standard-setting activity. We illustrate the sociological use of semi-structured databases by presenting our XML Schema for mailing-list warehousing. An XML Schema allows many adjunctions or crossings of data sources, without modifying existing data sets, while allowing possible structural evolution. We also show that the existence of hidden data implies increased complexity for traditional SQL users. XML content warehousing allows altogether exhaustive warehousing and recursive queries through contents, with far less dependence on the initial storage. We finally present the possibility of exporting the data stored in the warehouse to commonly-used advanced software devoted to sociological analysis.

## INTRODUCTION

Studying regular communication and public expression on the Web is becoming a common if not daily task for a growing number of sociological studies[1]. In this paper, we present the guidelines of a novel XML-based approach to deal with sociological Web data, such as the analysis of mailing lists or scientific publications available online. We thus propose new experimentation in the emerging field of computational social science (Lazer et al., 2009). Throughout this paper, we indicate in italics technical Computer Science (CS) concepts, which refer to a footnote the first time they appear in the article.

### The Particularities of Web Data

Though analysing Web data has become a major concern, managing this data is still done through very traditional methods which narrow the path of advanced research. Indeed, the vast majority of available studies in this field is based on manual extraction of data, and storage in simple *tables*[2] or *relational databases management systems* (R-DBMS)[3]. First, Web data is very heterogeneous. R-DBMS do not support the inherent heterogeneity of the various sources, and do not assist the scientist in the modelling and conception of sociological Web-oriented applications. These systems are poorly suited to social scientists' needs, and the internal structure of such databases is very complex and costly to apprehend for a external user. No modifications to the stored data model are in general possible. Secondly, the lack of adequate tools also exposes sociologists to the well-known risk of missing data, which can be caused, in the case of Web data, by the volatility of online publishing practices. Of course, the lack of exhaustiveness can be partially resolved by reducing selection biases. While such solutions are adequate when processing huge volumes of independent blogs (Drezner and Farrell, 2004; Adamic and Glance, 2005; King and Hopkins, 2007), what happens when the data to be stored is of continuous nature, and where sampling would diminish its significance (such as mailing lists), or if it is likely to disappear?

---

[1] Web data constitutes the primary source when one analyzes new kinds of socialization and civility (Beaudouin and Velkovska, 1999; Papacharissi, 2004), new forms of social movements (Diani, 2001; Pickerill, 2001; Van Aelst and Walgrave, 2004; Juris, 2004; Della Porta and Mosca, 2005 ; Calderaro, 2007), e-campaigning (Hacker et al., 1996; King and Hopkins, 2007 ; Drezner and Farrell, 2008), digital democracy and internet governance (Hague and Loader, 1999; Trechsel, 2007 ; Pavan and Diani, 2008).

[2] A table is the atomic entity of a relational database in which it stores information. Typically, an entity can be seen as a table, and each of its attributes is transformed into a column of this table. Each line in the table represents an entry in the database.

[3] Traditionally, since Codd (1970), one refers to Databases as Relational, or SQL (Structured Query Language) databases. Databases most used in the world are Oracle, IBM DB2, Microsoft SQL Server, Microsoft Access, or the freeware MySQL. Databases offer powerful querying facilities, implemented by the query language (SQL, XQuery).

Solving this kind of problem encourages the proposal of an implemented solution which would investigate the advanced potentialities of CS innovations for sociology and guide the sociological use of *semi-structured databases*[4] (Abiteboul, 1997) to bypass the technological rigidities of database schemas, and facilitate the interoperability of existing data analysis systems that use the same input data. In this paper, we advocate an XML-based solution for the management of sociological studies of Web data.

## Experimenting with Original XML Data Warehousing

The process of analysing online data can be broken into three distinct phases. First of all, there is the discovery and extraction of raw data. This means, for instance, finding a Web site containing data on a given mailing list that one wants to analyse. Secondly, this data must be transformed and stored in a local database or data warehouse. Finally, this data is processed by various applications, each one requiring a specific input format. By constructing, during the second phase, a generic warehouse of data, in XML format, which is standard, extensible, and easily transformable into any other format, we greatly improve the durability, usability and usage possibilities of the data (see Figure 12). Consider for instance in France, some research programs such as Marlowe (Chateauraynaud, 2003). Its objective is to ease the extraction of Web data and to manage pragmatic frameworks for the study of debates and controversies on the Web. Indeed, the Marlowe program has provided significant advances in the field, mostly focused on easing user interaction with the system, by the use of natural language communication, and specific heuristics to understand natural language. However, the documents studied by Marlowe (stored in Prospero, or news feeds acquired by Tiresias) are not stored in a format that facilitates cross analysis or partial extraction by another system. Indeed, these internal structures are focused on the objectives to be fulfilled by Marlowe.

We advocate the use of an intermediary storage system, generic and exhaustive from the raw information point of view, which can expose this data to any software that wishes to process it (such as Marlowe). Consider for instance the interesting study of certain collaborative work on the Web (Dorat et al.; 2007; Conein and Latapy, 2008). The authors extracted raw data from mailing lists managed by open source communities, using programs written in the Python language. Their data covers messages posted by 2,287 different email addresses, and the authors consider that each email address corresponds to one individual, but recognise that this is not true in practise and that this has little influence on the results. A different task, that we undertook in our case study on the XQuery mailing list, would be to determine to which real individuals the emails correspond, and to construct the correspondence between a person and a list of emails, possibly with dates of usage. By storing such a result in an XML warehouse, rather than using a proprietary file format, it is very easy to analyse the list from the posted emails point of view, or the "physical individual" point of view, thereby facilitating the interaction and interconnection of studies on the same dataset that could be carried out concurrently. Therefore, we stress that the approaches adopted by both

[4] Semi-Structured or Native XML Databases are still hot research topics in Database Technology. The difference between a relational and an XML Database is the format of information that can be stored. In an XML Database, information is far more flexible, which leads us to prefer this type of data store for sociological applications, where the schema is difficult to define once and for all.

Chateauraynaud, and Conein and Latapy present innovative data processing, but unfortunately low interoperability of results or original data, without a good knowledge of the internal working of the system. We advocate the use of an intermediate XML data warehouse to store the data, or even some processing of results.

There is actually a new research trend with new interdisciplinary programmes mixing sociologists and computer scientists who propose advanced XML solutions. In the United States, for example, the Data Documentation Initiative[5] (DDI) has developed XML solutions which converge with the one we experimented: the XML format is viewed as the ultimate standard to produce, store and use data in an interoperable way. To interface with a system such as Marlowe, we simply need to write a transformation program in a dedicated language (XQuery) which extracts the relevant XML data in Marlowe format.

As we have suggested, the use of an XML warehouse to store data is inescapable. It is however possible to go even further and to process XML data using the XQuery language to analyse both structure and content of the data. To illustrate this possibility, we present experimentation based on the case study of the standardization process of the Web in the World Wide Web Consortium (W3C), which was historically our first application. In the particular case of elaborating Web standards, mailing lists are all the more interesting since they are becoming the prevalent means of interaction between participants scattered around the globe and working in different time zones.

Our data corpus from 8 public mailing lists of the W3C, which have been active since 1999, was extracted automatically. In addition, we also extracted from the W3C Web site an inventory of the technical preconizations[6] produced in connection with these lists, and from DBLP[7], an XML database, all the bibliographical data concerning our community of CS experts. Here, we present some exploratory steps of the case study from which our model was elaborated as a guideline for future users of XML data warehouses, and an invitation to structure cooperative work in this field.

In this paper, we first present XML and how to model a data warehouse in the next section. We then introduce the management of data sources in the third section, before examining practical sociological enquiries in the fourth section, and discussing in the last section the added value of such databases for sociologists, and further methodological challenges they will have to face.

## BUILDING AN XML DATA WAREHOUSE

The use of a data-centric application can be divided into four different actions that are not necessarily sequential: constructing the model of the warehouse, loading the data into the warehouse, and querying the data. A fourth optional action is possible: the conversion and export of this data in a different format to a third party application. In this section, we focus on the first aspect: building the schema of the warehouse.

### The XML Data Model

The XML data model is merely a tree[8], where each node of the tree is of a specific type. There are a dozen different node types in XML. In this article we will only consider the three main types:

---

5 http://www.ddialliance.org/

6 As technical preconizations, we include: official recommendations of W3C, drafts, which are supposed to become such recommendations, and Working Group notes.

7 Available at http://www.informatik.uni-trier.de/~ley/db/ DBLP is a well known database on computer science publications.

8 A tree is an acyclic connected graph (see http://en.wikipedia.org/wiki/Graph_%28mathematics%29).

element nodes, attribute nodes, and data nodes. Element nodes have a name – in our example, these are <actors>,<actor> and <name> – and can have any number of child nodes. A data node is comprised solely of data (numeric, text, or binary data) and has no child nodes. An attribute node has both a name and a value, and has no child nodes either. Note that attributes can always be replaced by elements[9]. Element and attribute names can roughly be any sort of text, or typed data, such as in integer or decimal form. By giving names and values to these elements, one can build an XML document instance (simply called XML document).

**Insert Figure 1 here**

Elements are delimited by opening and closing tags. An opening tag is composed of the name of the element in between bracket, and a closing tag is composed of the same name between bracket, preceded by a / symbol. If there is anything in between the opening and closing tags, then this means that the element is composed of other XML nodes. In our example, the <actors> node is composed of two <actor> nodes, each of them is composed of a <name> node, containing some text, and an attribute called *id*. The values of attributes are between inverted comas. As you can see, contrary to SQL, it is possible to construct an XML document without having to defined any kind of structure[10].

## Conceptual Modelling

Writing XML documents is easy, but it becomes very difficult to write a document that will contain all the information needed without spending some time building the *conceptual model* of the information that will be stored in the database.

Databases are constituted by fundamental abstract elements which are as many categories as needed and have a logical signification, usually *entities* and *relationships* (Chen, 1976)[11]. An entity represents a class of objects, individuals, concepts, etc. (i.e., we could construct an *actor* entity¸ defined by a *name* element and an *id* attribute). A specific actor element (i.e. Michael Kay) is called an instance of an entity. A relationship links two different entities. For instance, the fact that an <actor> <works for> a <company> is a relationship. We will discuss relationships in more detail in the advanced features paragraph below. To summarize, an entity instance is necessarily an element, but an element is not necessarily an entity. Attributes simply characterize entities.

The meaning of the data is constructed by the sociologist when conceiving the warehouse schema (i.e., when defining the entities and relationships), which is called conceptual modelling. Of course, the meaning can be enriched when interpreting the data and writing queries (i.e., we can add a concept of <top poster> for an <actor> who posted over 100 messages). Therefore, the conceptual model of a data warehouse (i.e., the entities and relationships it represents) depends on the categories and background hypotheses constructed by the sociologist. Nevertheless, two different XML conceptual models can be integrated into the same query.

---

[9] An attribute name of a given element must be unique, whereas this is not the case of elements. For instance, if email is an attribute of <person>, then a person can only have one email value. To have several emails, <email> must be an element.

[10] To enter data into an SQL database, one must first create the database schema (i.e., define which tables are to be used in the database).

[11] These refer to the Entity-Relationship model, used since the 1970s to model relational databases. This model professes that any object in the world can be abstracted by either an entity (if it is an independent object) or a relationship (if it only exists through other objects). Attributes are the atomic characteristics of entities or relationships. This model can be applied to both relational or XML databases.

Each document, such as the document shown in Figure 1 is called an instance of a given schema if its structure is valid[12] with regards to this schema. An interesting characteristic of XML documents, which leads to better extensibility, is the fact that a document can be valid with regard to several different schemas. We detail below the model that is actually used by our application, but let us first comment on this introductory model.

Figure 2 shows a graphic representation of the schema, based on the commercial Altova XML-Spy tool. We will be using this tool to represent our schemas throughout this article.

**Insert Figure 2 here**

Elements are shown in rectangles, and octagons represent a sequence of elements on their right hand side; the cardinality of this sequence is indicated below the octagon and, by default, cardinality is 1. For instance in our example, an <actors> element can have a minimum of 0 and an unbounded number of <actor> elements below it. An <actor> element has at least and at most a single <name> element. The lines on the top left corner of the <name> element indicate that it can have some text below it. One can easily verify that the document shown in Figure 2 conforms to the schema of Figure 1. To show that a document is valid with regards to a given schema when writing the document, we can refer to this schema by creating a namespace which is a reference to the schema that is used as a prefix to the element names. The reference to the schema is in general a URL pointing to the schema, written using the XML-Schema syntax. Suppose that our schema is at the location http://www.prism.uvsq.fr/~beng/WebStand/Schemas/simpleActors.xsd, we show in Figure 3 our initial example document using a schema and a namespace, which is done by adding a special *xmlns* attribute, that indicates the prefix, **act** in our present case, that refers to a schema. To be precise, an XML schema is defined by an XML document using the XML-Schema namespace, and we show its definition on the right of Figure 3. We hope that understanding this XML document is by now pretty straightforward.

**Insert Figure 3 here**

## Advanced Schema Features

There are of course many interesting features in an XML Schema that we do not detail here. However, we will point out the referencing feature which models relationships used to perform joins; that is, to merge information from several documents that are talking about the same entities.

In SQL, it is common to define keys in tables so as to uniquely identify an entry. For instance, one could assign a unique combination of letters and numbers to each physical person, called a primary key, and use this number in other tables where it is called a foreign key, to show that the information is related to this specific person. The same is true in an XML Schema. It is possible to define a given attribute or element as a key by giving it the type xs:ID. Foreign keys are defined using the type xs:IDREF. There is always a big difficulty in sociological databases to correctly construct a key. We use here an alpha-numeric key, mainly based on the *firstname* and *lastname* of the individual.

---

[12] We say that an XML document *d* is valid with regard to an XML Schema *s* if the names and structures found in *d* follow the constraints defined in *s*. We call a type of an XML element the structure it must obey.

**Insert Figure 4 and 5 here**

Figure 5 is a document that represents Michael Kay (*id*="A1") as CEO of Saxonica (*id*="I1"). This document is valid with regards to the schema of Figure 4.

## THE MAILING LIST APPLICATION SCHEMAS

Any social scientist who wishes to construct a database schema has to elaborate a conceptual sociological model that fits with database management science. We feel that this task is achievable by the sociologist.

Unfortunately, translating a conceptual model into a concrete database schema requires technical computer programming skills that few sociologists have already acquired: They need help from computer scientists. We have tested a general model of conception which takes into account the extended universe of sociological constructs (from entities such as persons, institutions, etc.). Future work involves using this methodology as the basis of a system, composed of such bricks that researchers could modify at will. We have currently developed several bricks, such as schema construction, CV extraction, etc., but some implementation work still remains.

### Sociology Oriented XML Schema Construction Methodology

Our methodology can be divided into the following phases:

• 1. Find the high level, raw data entity. An entity is an object that has some kind of real or virtual existence. Any entity that is the result of a calculation, for instance "top poster" does not fall under this category.

• 2. Find the high level relationships between these entities. One major difficulty in conceptual modelling is to define what is an entity and what is a relationship. In our approach, we model as relationships abstract concepts that link two entities. For instance a <car> would not be a relationship between <engine>, <wheel>, etc. since it is itself a physical entity.

• 3. Define the compulsory characteristics of the entities and relationships. Each characteristic must be an attribute, if it is invariant and mono-valued throughout time, or an element in all other cases. For instance, the *id* of an actor will be an attribute, possibly his *dateOfBirth* also, but his <name> will be an element, since it has a complex structure composed of a compulsory <lastname> and other optional and multivalued elements such as <firstname> and <middlenames>. Note that phase 3 can be done in parallel for all the entities and relationships.

• 4. Check coherence of phase 3 by verifying that raw information is only present in a single entity, and that references are used in all other cases. In what follows, we present the results of this modelling phase in the form of several XML Schemas and, as we will see later, customizable queries.

### Determining the High Level Concepts

If a database concerning people interacting on forums is constructed without defining entities such as <mail> or <actor>, it would simply be impossible to measure anything about individuals who posted mails. The choice of categories, which means a sharp definition of them, is crucial for the success of the study. These preliminary remarks seem trivial, but represent a major challenge when trying to conceive a good schema. When one constructs an entity, one postulates that its essential (i.e., always present) characteristics will be specified in the database (i.e., the <name> of an <actor>). Some other characteristics may not be, but they will become secondary ones, so that they will not help (initially) to establish relations between entities of the database. Since we also want to work on professional trajectories, at least three entities will be needed: <actor> and <institution>, that are the main entities, and a relationship entity – <function> – whose action is to link an <actor> to an <institution>, as shown in Figure 4. This is one illustration of the power of using an XML warehouse: joining data over multiple dimensions and multiple data sources. Each data source might even have been constructed by a different organisation!

The other side of our model deals with the representation of messages on the mailing list: an email discussion is composed of many <message>s that can be regrouped into a <thread>, that has an initiator message, and then messages that answer each other, as illustrated in Figure 6: M1 and M3 are messages that create a new thread, M2 and M4 are both messages that answer M1.

**Insert Figure 6 here**

## Determining the Relationships between Entities

There are several relationships that appear in the model. The first is the <institution> that each <actor> is part of, and the role played by the <actor>. We have already illustrated this in the section above. The other main reference problem of this application is the link between a <message> and an <actor>.

As a matter of fact, we chose not to store a given <actor> as author of the <message>, but instead an <email>. It is then possible to define which <actor> posted a given message by computing a join on the <email> value by assuming that an email uniquely identifies an <actor> entity. The reason for our choice is that we regroup all the emails that we believe belong to a given person, but if at some point we change our mind, the reference data is left unchanged. Let us illustrate this with the following (real) example. Don Chamberlin, an important IBM poster had been identified, using his email don@us.ibm.com. We also had another poster called xquery@us.ibm.com who was a priori a different actor. However, by analysing automatically the contents of the emails, using a simple XQuery program, it transpired that both were the same person. We therefore only had to merge the <emails> elements of both <actor>s to achieve this identification. The rest of the data didn't need to be changed.

Moreover, since an email address contains interesting information, such as the institution sending the email, by storing this specific information, it is possible to determine via which <institution> an <actor> is posting.

## Determining the Characteristics of the Entities

In our study, we found that an <actor> entity only had two primary characteristics: at least one <email>, and a <name> composed of at least a <lastname>. The actor entity also had many (optional) secondary characteristics such as <firstname>, <middlename>, <sex>, <birthDate>, a list of <diploma>, <skill>, etc. This information can be made present as optional elements or attributes in a flexible manner. We stress that other optional entities can be added without causing any trouble with existing data or queries. Similar to Figure 4, we use a reference attribute of type xs:ID as *primary key*[13] to the individual. Since in our context, all of the actors have names, no anonymous individual could take place in cross-company comparisons, except if complementary models are added to bypass the lack of information (Jansen et al., 2006). To do this, we would have to relax the constraint of our schema that says that all individuals have names, and instead say that this characteristic is optional.

Concerning the message entity, apart from the characteristics already discussed, such as the message ID and the email address from which it was posted, we also store other information, such as the full textual content of the email, the date (although this could also be done using our temporal model), the subject, as well as the identified topics of the message. In the first model, we considered each <email> was unique, and associated to a single <actor>, by setting the email as an xs:ID. Although we have no such example in the XML/XQuery data set, it is nevertheless possible for several <actor>s to use the same <email>. To deal with this, it would be necessary to add a link to both the <actor> and the <email> used when posting a message.

### The Schema

The definition of the two following schemas, following our methodology, was elaborated using both our in-house tool and the commercial XMLSpy software. We represent, on one hand, the posters and their affiliations, with the actors_info.xsd schema (Figures 7 to 10) and, on the other, their messages, using the threads.xsd schema (Figure 11).

**Insert Figure 7 to 11 here**

Figures 7 through 11 show the XML Schema of the database with its three main entities (implemented as elements) <actor>, <institution>, and <function>, complete with attributes and the <threads> structure. This schema can be extended at any time by adding new elements or attributes, and all queries already written using the existing schema will continue to function with the new schema. This is a major improvement over SQL where all the queries need to be checked for compatibility, in the case of schema evolution.

On the basis of this experimentation, other researchers studying mailing lists can either directly use our schema; or construct their own one according to the instructions presented in the previous subsections. The most specific element is the recursive <message> element. Such a structure can not be built in basic SQL, and is in any case complicated to compute, even using SQL extensions. Moreover, the flexibility of an XML Schema allows many adjunctions or crossings, which are very useful and simplify research work. For example, we started to study our data in a synchronous manner before integrating a new temporal sourcing model (see Colazzo et al., 2008) related to entities and attributes such as emails, functions, or dates of publication of standardization drafts as sourced facts.

## MANAGING DATA SOURCES: AUTOMATIC EXTRACTION / CLEANING / MANUAL VALIDATION / ENRICHMENT

[13] A primary key is a set of attributes that identify in a unique manner an entity. In the real world, the couple FirstName/LastName is *not* a primary key of the entity [person], since there can exist two different people that have the same firstname and lastname. However, in the context of people in the W3C arenas, this seems to be the case.

To extract and store large quantities of information, it is critical to design a methodology which needs as little human intervention as possible. However, human input and feedback can (and should) be used to adjust and enrich the system. When extracting data from mailing lists, our process model is organized as follows:

**Insert Figure 12 here**

As shown in Figure 12, we are interested in identifying actors: the individuals that post messages on the mailing list. We have presented the model used to store information. We map data sources of interest to entities and relationships of this model and load the data into our warehouse (Dudouet et al., 2005).

A number of practical issues require the use of automatic and semi-automatic filtering and enrichment: The name of an institution may appear written in many different ways. We use classical text mining techniques to establish that "Sun Microsystems Inc." is the same institution as "Sun" and "Massachusetts Institute of Technology" is the same as "MIT". Further identification can be done manually (i.e., to understand that "cerisent.com" is a Web site belonging to "Mark Logic Corporation"). A person can have several different email addresses; identification of persons must thus be done on their names and not on their addresses (assuming that two persons do not have the same name). In a relatively small group (involving about 200 people), a simple manual check, based on automatic extraction of names from email addresses, suffices to make sure two actors do not share the same firstname/lastname pair (which we found to be indeed the case in our study). For a larger group, a data cleaning tool on a person databases could be applied: currently, there is some research being done in the database community to find simple ways of defining if two homonyms are the same person or not, based in general on their links to other entities (Kalashnikov et al., 2007).

In our case, first name, optional middle name, and last name are extracted from the full name description making possible the assimilation of "FirstName LastName" and "LastName FirstName" patterns for instance. We have not yet introduced any data cleaning based on structural relations.

The content of these two warehouses is what we call "raw" data; we may enrich it manually with extra information on actors and institutions. The data extracted automatically can be complemented by other information sources found for instance on the World Wide Web (HTML or XML data describing mailing list posters, home pages of the more important actors institutions Web sites) using wrappers. We have used various information retrieval processes to help with such work. For instance, to efficiently find Web pages of an actor or organization, we can exploit some techniques for identifying relevant pages by using some non-content features, like page length and the URL form (Kraaij, Westerveld and Hiemstra, 2002). In addition, we can exploit other techniques based on the use of existing Web search tools (e.g., Google) to find a coarse list of potentially relevant pages, to advise on the use of particular information retrieval techniques able to extract relevant and representative information contained in these pages, and to guide the user in the identification of searched Web pages (White, Jose, and Ruthven, 2001). It is of course technically possible to combine automatic fetching (extracting raw data) and information retrieval (screening relevant data) from available data online. In any case, our procedures will always be semi-automatic, which means we do not aim tot create humanoid robots such as automatic sociologists or something else. Our prototype generates propositions which need to be manually confirmed. We refer to Pinchedez (2007) for more information on our prototype, based on the Exalead search engine.

## IMPLEMENTING SOCIOLOGICAL ENQUIRIES

Now that we have explained conceptual modelling, data cleaning and enrichment methods, we can take a quick look, with our case study, at the kind of work sociologists may implement with a XML data warehouse.

Our sociological study of the W3C is part of a larger study of the international standardisation process of ICT, in which we investigate how standards are established at an international level. In our research, we investigate the structure of interactions on mailing lists, and provide a dynamic view of the figuration of expertise. Quantitative analysis consists mainly of counting the number of emails sent on mailing lists by individual actors, actors from a given institution, etc. This gives initial indications on the two levels of personal and organisational investments in this activity. We will first show how to obtain this basic information with simple XQuery examples. Then, we will present how we went back from mailers to the institutions (firms, research centres, NGOs, etc.) to which the actors were linked. Our aim was to identify which institutions were the most implicated in the standardization process. In this section, we also illustrate how XQuery (Deutsch et al., 1999; Chamberlin, 2003; Fernandez, 2004) can be used to integrate other XML databases, such as the W3C list of author recommendations.

### Exploring the Warehouse Using XQueries

Automatic extraction and well-structured warehousing of course does not mean that one gets a ready-made data base. Implementing direct queries on the stored data is a first step of exploration, but does not automatically provide sharp data, as we will show below. Since we do not want to overload this paper with technical code, but simply show what sort of queries it is possible to execute, we have not included the code of the queries here. However, it is available online at the WebStand site, along with the "raw" XML result document[14]. We refer to the queries by their code, $BMS_{Q1}$, $BMS_{Q2}$, etc. to simplify referencing them on our Web site.

The first range of inquireis consisted of measuring activism on these 8 lists. We started by computing the number of emails posted per person. Let us stress that this query gives different results than simply counting the number of posts per email address since many of the important posters use several email addresses: Michael Kay uses 3 different ones, Ashok Malhothra uses 6, Dana Florescu uses 4, etc. As a matter of fact, the longer the period of posting goes on, the more email addresses a person will tend to have, since most posters use their professional email, and therefore it changes when they change institutions.

Query $BMS_{Q1}$ computes the names of the actors and number of posts for the people having posted at least 20 messages (this value is of course a parameter that can be changed). This gave us 72 actors (out of a total of 334) who sent out a total 10,619 messages, which represents 61 percent of the total interactions, if we exclude the mails sent out by a so-called Bugzilla (3944 messages). Defining the value for which a poster will be considered important is of course the sociologist's work. This can be done for instance by studying the distribution of the number of messages posted, and choosing a cut-off point.

This first count of the messages sent by authors indicated that some people we had previously identified as influential, because they had signed recommendations, were absent from the warehouse! On another hand, we identified with this count this strange author named Bugzilla as a notification system at first glance. By exploring the content of Bugzilla's messages, we understood this author's name was a generic one given to an automatic inner bug reporter of the mailing list. The real email used to send this information was in fact hidden inside the core text message sent by Bugzilla. Retrieving the people we had identified obliged us to write new full-text queries (i.e., queries that analyse the text contents of messages) to extract their messages from Bugzilla and obtain the right measurement of posters and mails by doing so. XML content warehousing was very useful here because only the query counting the messages needed to be rewritten. The whole extraction process was only executed once, and the raw data stored in the warehouse remained unchanged.

[14] http://www.prism.uvsq.fr/~beng/wiki/index.php/WebStand#BMS_paper_queries

In our exploratory queries, we also identified which posters were multi-positioned (i.e., had posted on several different lists; see $BMS_{Q2}$). The interesting result here is that nearly all the important posters were multi-posters. Only one multi-poster that had not sent a large number of emails to the lists was Tim Berners-Lee[15] himself, who is of course someone noteworthy!

We now investigate temporal queries on a single list (it is of course possible to compute the same results on several lists). We chose to study the public-qt-comments list on the XQuery recommendations. In $BMS_{Q3}$, we compute the total number of posts issued per month. This is a simple aggregate query, and of course any time-lapse or "granularity" can be chosen (day, year, etc.).

**Inser figure 13 here**

We see that there is a very important number of posts in 2004-02 (1251 messages). Once again by using the full text capacities of XQuery (see BMSQ4), we discover that most of the titles of the messages contain "Last Call Comments on XQuery 1.0". Indeed, February 2004 was the deadline for the Last Call Comments action of the W3C, which is basically the last moment when it is possible to raise objections against the specification without stopping the whole process, and hoping that these features will be included. Such important activity on the mailing list can be justified by external information/knowledge, but such information can also be discovered through a full text query.

In query $BMS_{Q5}$, we select all the emails of the <actor> named "Kay". For each of these emails, we compute for each month the number of actual messages posted by this email. As you see in this example, we use XQuery built-in operators to cast a full ISO Date to the gYearMonth type. The results of such a query are better viewed graphically, as in Figure 14, where we plot on a common timeline the number of messages posted from each of his email addresses. As you can see, there is a clear cut in 2004-02, where the address Michael.Kay@softwareag.com (in white) is no longer used, and mhk@mhk.me.uk (in black) begins getting used, and another cut in 2005-04 where the Bugzilla system was introduced and the actor started posting via this system (hashed bars). The first cut corresponds to a change of institutions, while the second corresponds to a change of posting methodology to the list. Also note that this poster had a high participation in the Last Call Comments in 2004-02 not as an initial poster (discussion starter), but as a respondent. Indeed, an analysis of his signature shows that he signs as XSLT editor, which explains why he is responding so actively to all the comments. Note that this month also corresponded to his change in affiliation. Investigating if this is a coincidence could be done by writing more full text queries to gain some information on the contents of the messages posted around the date 2004-02.

**Insert Figure14 here**

[15] The inventor of the Web, and Director of W3C.

To conclude this section, let us stress that as shown above, the result of an XQuery is itself an XML document, that can therefore also be queried. Our experience has shown that sociologists like to test various queries, and reuse some results to construct more complex queries. To simplify the reuse of intermediate results as sources for new queries, we devised a graphical query system integrated in the WebStand platform, demonstrated in Nguyen et al. (2009). This system also aids neophyte users to construct basic XQueries using a graphical interface. This system is available as a free download[16].

We also feel that an important result of our study is a library of customizable XQueries that can be used to analyse mailing lists, provided that they are stored using the XML Schema we proposed, or a customized extended version of our schema. Thus the source code of all our queries is available on demand, and can be used freely.

## Dealing with Affiliation: The Added Value of XML Warehousing with Hidden Data and Crossing Data

### *From Actors to Institutions*

As presented in the section above, the existence of hidden data, such as Bugzilla's posting activity, engenders major complexities for SQL users, since the retrieval of the real messages authors' identities was not a task initially planned. XML content warehousing simplifies the management of this kind of hidden data. Integration of other online XML databases is also very simple; as a matter of fact, it is even sometimes transparent for the user.

In our case study, we decided to observe the concrete activity of experts of the W3C and then look for their ties with companies or other institutions concerned with innovation. Such a method leads to measures for detecting the most active institutions in this process. The following table is the result of query $BMS_{Q6}$.

**Insert Table 1 here**

As shows, firms are very active: people from Microsoft, Software AG, IBM, Oracle, Saxonica, have posted over 1,000 messages. We had to be cautious with this measure, because some individuals might had successive memberships.

When jumping from actors to institutions, one faces two kinds of errors. The first sort of errors is related to the temporal limits of the functions assumed by the actors. A given actor may be linked to several functions which indicate his professional trajectory. A synchronous view of these functions from the simple relationship between the actor's id and the email addresses implies fake multi-positions: these are successive positions. In our corpus, many computer scientists moved from one company to another during the time period considered, without actually being multi-positioned (i.e., belonging to several institutions at the same time). This professional mobility is not trivially derived from the temporal succession of email addresses. As a majority of email users (Buckner and Gilham, 2000), people often use personal addresses during the transition phase, and in consequence their belonging to one company or another is fuzzy. This is why functions must be considered as sourced facts and dated chronologically as such (see Nguyen et al., 2010).

---

[16] Available at http://cassiopee.prism.usvq.fr/WebStand/

*Hidden Data*

The second kind of error is related to what we called hidden data. As we explained, our corpus contained about 4,000 mails sent by some mysterious “Bugzilla”. Using XML full-text content querying, there is a basic operation of XML content warehousing which is an efficient way to identify the actors who sent mails using the Bugzilla mailing system. As said earlier, it was possible to keep the raw data and to pursue deeper investigations on message authors by exploring one step beyond the content of the emails. In this case, a poster could have two typical bug reporting functionalities. The first one was marked as “reported by”: this indicated the poster was reporting a bug from their experimentation of the technology which was being standardised. The second one was marked as “comments from”: this indicated someone was taking into account the bug report and was either proposing a ready-made solution or filling the “to do” list. From this information, we just had to write more pointed queries which would bring out email addresses and dates from the emails marked “reported by” or “comment from” (this explains the two different bars for the emails sent by Kay using Bugzilla, as shown in the results of $BMS_{Q5}$ in Figure 14). In this case, the information was directly available, but we could have also have analyzed another part of the contents of the message, such as the signature. XML content warehousing thus allowed us to restore the correct authors of messages in the database without modifying the raw data through manual corrections, which might have been necessary in SQL. This is of particular importance in the case of mailing lists where one can find inner posters or more generally posters who send from friends’ mailers. The limit of this methodology is that social scientists should always respect privacy rules in their investigation. In our case study, as mailing lists were published online and Bugzilla was a public mailer, there was no risk of breaking privacy rules.

*Integrating Other Online Databases*

Another added value of XML content warehousing is the simple integratioin of other XML databases. In our case study, we were trying to establish a correlation between the actors’ activity on the mailing list and their influence on final recommendations, using indicators such as co-authoring. From the contents of the recommendations edited online, we first implemented an extraction of the relevant information about the co-authoring of the edited recommendations linked to the lists we were studying. Later, we were informed that W3C had already stored comparable data online[17] in RDF format, which is compatible with XML. Using this new database, we were able to cross-check our data. We only had to establish new relationships between email authors in our database and recommendation authors in W3C’s database (see BMSQ7). Such integration, which is very useful in any sociological enquiry, is much more demanding with SQL databases. As a matter of fact, XML is commonly used to integrate SQL databases, but one has to write a wrapper application. XML’s structure is much more flexible for any initiative of data-crossing.

We illustrate such cross-data source analysis by crossing the important posters (see $BMS_{Q1}$) on the W3C list with the DBLP XML database, illustrated by the results of $BMS_{Q8}$ shown below (figure is normalized so that the top poster (Michael Kay) has a posting of 100 percent and top publisher (Dana Florescu) has a total number of articles of 100 percent) (Figure 15).

**Insert figure 15 here**

[17] http://www.w3.org/2002/01/tr-automation/tr.rdf

In this case, we queried the DBLP database with the important posters found in $BMS_{Q1}$. $BMS_{Q8}$ counts their total number of publications, and compared it with their total number of posts on the lists. We chose here to parameter $BMS_{Q1}$ with 40 posts, which gave us 27 posters, of which 20 had published at least 1 article in a scientific conference. This gives a global proportion of 74 percent of publishers amongst top posters, which can be compared to results of query $BMS_{Q9}$ for all the other people on the lists of which only 32 percent have published at least one article. It is interesting to note that among the small posters who have published articles, all of the important posters were from academia (and only 3 of them had posted more than the top poster of the top publisher sub-group). On an average, top posters have a proportion of 11.33 articles per person, whereas low posters have an average of only 5.7. All this information converges towards our conclusion that quantitative analysis of posting habits provides exploratory results for profiling experts in such a standard setting arena.

*Exporting Data to Common Software Used by Social Scientists*

As we have already discussed above, one of the interesting capacities of XML is to be easily exportable to other applications, or to other application formats. For instance, it is possible to export XML temporal analysis data in TraMineR, and the transformation into R format used by TraMineR can be done either before by an XQuery, or in R itself by using existing R-XML packages. Prospero/Marlowe use their own data format, so to have data processed by such a system, we simply need to write a query to transform our data into that format. This is a general advantage of using an XML data warehouse: the raw data is in a single place, therefore there are limited problems of inconsistency. Whenever data is changed, then it needs only to be changed in the XML warehouse. The values used by other systems can then be regenerated using the same query. This whole approach can in fact be pipelined for more efficiency.

## DISCUSSION AND CONCLUSION: XML CONTENT WAREHOUSING, A STEP BEYOND SQL POTENTIALITIES

*Flexible Data Wharehousing*

First, we have shown how the use of state of the art developments in XML database technology can lead sociologists to be able to apprehend vast quantities of data readily available on the Web. Many XML databases are available online, since XML is currently viewed as the best standard for publishing information on the Web. Building XML warehouses is here the most flexible solution to store and exploit the data. An XML Schema allows many adjunctions or crossings, without reprogramming sets of data as soon as entries are different. As shown in the example of database crossing, the tree pattern structure is very useful in implementing queries without reintegrating data in the same schema. This is much more difficult to achieve in SQL. Moreover, XML can integrate SQL data by the simple use of foreign keys, using the type XS:IDREF. As a result, XML content warehousing is also one of the best ways to manage peer-to-peer databases, which are very useful for social studies, as the enrichment of data is a constant need.

*Exhaustive and Recursive Data*

As mentioned earlier, XML content warehousing is very useful in exploiting continuous data like mailing lists, which are very difficult to sample. Content sampling with bias reduction through Bayesian bootstrap analysis or alternative methods is achievable when non-exhaustive contents do not alter the comprehension of what is at stake, especially when samples of contents are related to information or comments based on a same sourced fact, like in buzz processes. As soon as exhaustive contents are needed, warehousing methods have heavy implications on further querying. In our schema, for example, we have shown that the recursive <message> element (see Figure 8) can not be built in basic SQL, and is in any case complicated to compute, even using SQL extensions. We have also shown that the existence of what we called hidden data implies major complexities for SQL users. XML content warehousing allows altogether exhaustive warehousing (what is also achievable in SQL) and recursive queries through contents, with far less dependence on the initial storage.

*Available Data for Mix Methods*

As shown in a section above, managing exhaustive and recursive Web data is not only a way to go back to this data with new ways of categorizing its basic entities or new needs for the exploration of their content: it is also a good way to do this while going further by using mix methods. In our case, for example, profiling the experts of a W3C working group can be achieved by crossing their activity in collaborative work (from our original extraction) and their academic activity (from the bibliographical data automatically imported from the DBLP database); then, all this data can be exported very easily into both software devoted to sequence analysis and optimal matching (such as TraMineR), and software conceived for content analysis (such as Marlowe): in this case, mixing studies of academic careers and studies of controversies would help in understanding expertise from both the dynamics of experts' scientific experience and the dynamics of their argumentative practices. Managing an XML warehouse helps doing this without collecting and restructuring the whole data at each specific step of empirical investigation, which means it helps to manage collaborative and cumulative work from data stored in a given format once and for all, and to centralise editing of updates if necessary.

We expect to investigate this further on still larger samples, using analogous techniques. We also expect to establish creative cooperation with teams engaged in comparable experimentation on various mailing lists.

## WEB SITES

.

FIGURES

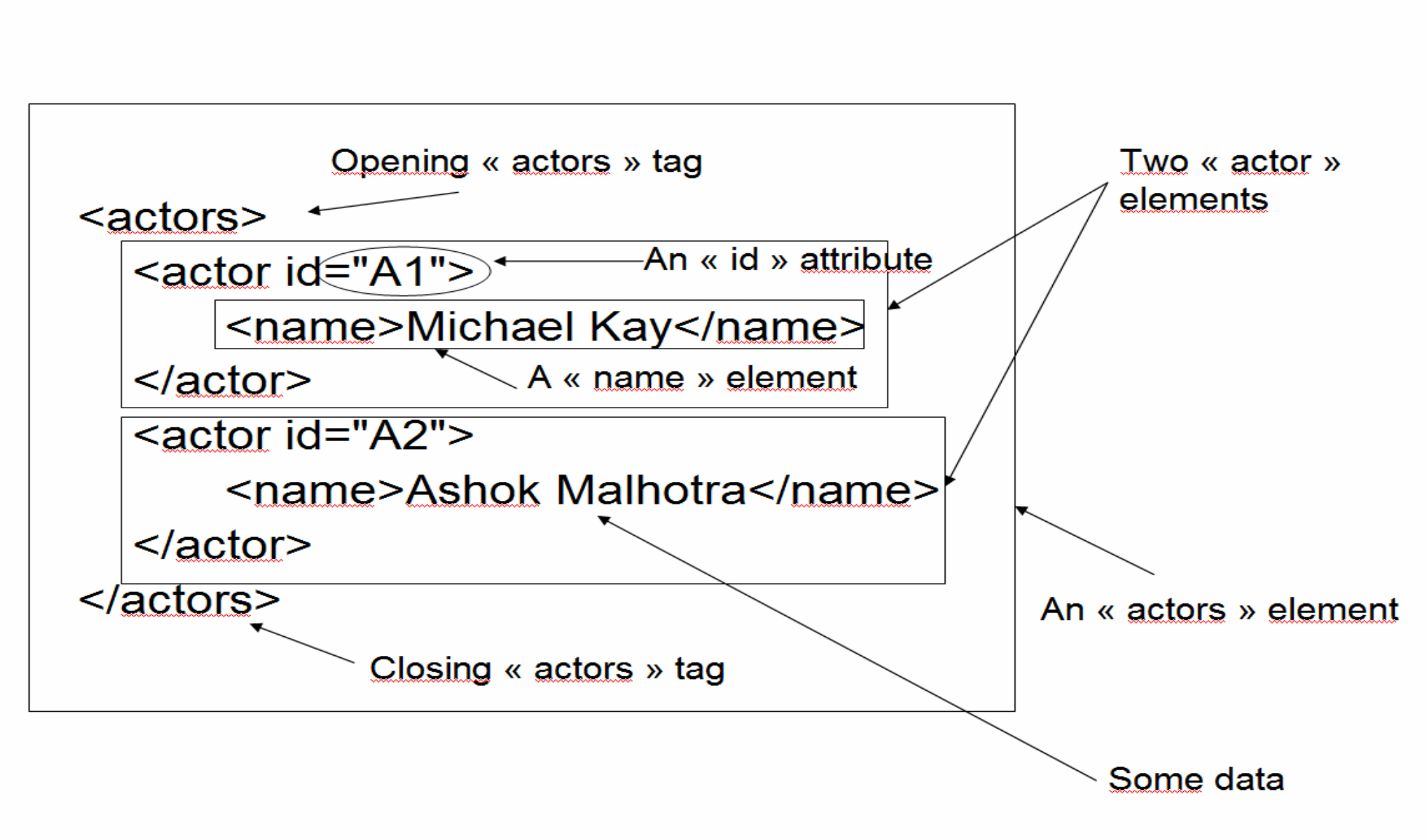

**Figure 1 -The text version of the *actorsExample.xml* document**

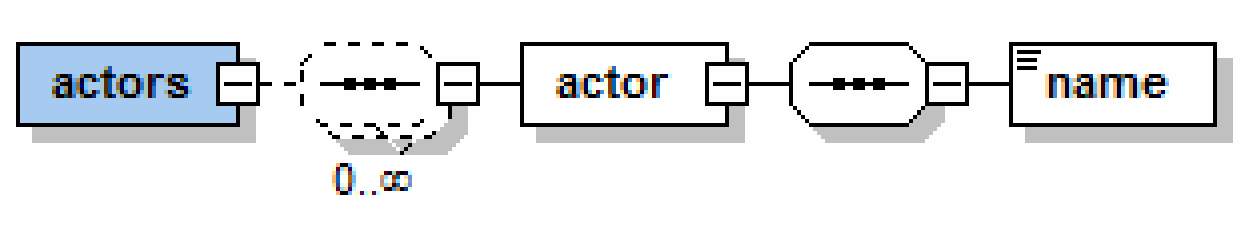

**Figure 2 - XML Spy representation of the simple actors schema**

```
<act:actors
xmlns:act="http://www.prism.uvsq.fr/~beng/
WebStand/Schemas/simpleActors.xsd" >
        <act:actor>
         <act:name>Michael Kay</act:name>
        </act:actor>
        <act:actor>
         <act:name>Ashok
Malhotra</act:name>
       </act:actor>
      </act:actors>
```

**New *actorsExample.xml* document using a schema**

```
<xs:schema
xmlns:xs="http://www.w3.org/2001/XMLSche
ma" >
       <xs:element name="actors">
        <xs:complexType>
         <xs:sequence minOccurs="0"
maxOccurs="unbounded">
          <xs:element name="actor">
           <xs:complexType>
            <xs:sequence>
             <xs:element name="name"/>
            </xs:sequence>
           </xs:complexType>
          </xs:element>
         </xs:sequence>
        </xs:complexType>
       </xs:element>
      </xs:schema>
```

**The simple actors schema in XML**

**Figure 3 – Making an XML Document of actors through an XML Schema**

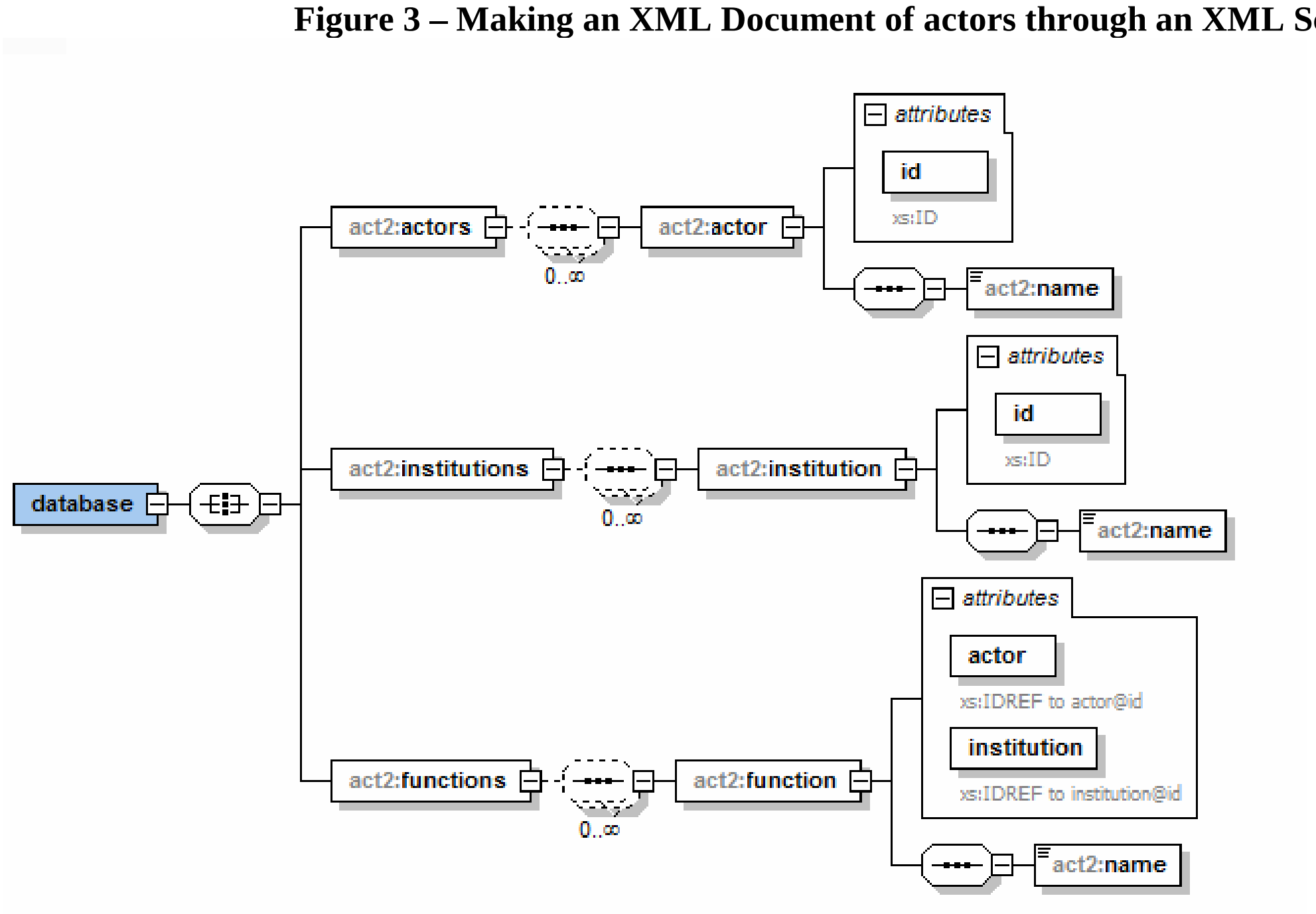

**Figure 4 - Schema using a relationship**

```
<act2:database
xmlns:act2="http://www.prism.uvsq.fr/~beng/WebStand/Schemas/extendedSchema.xsd" >
 <act2:actors>
  <act2:actor id="A1">
   <act2:name>Michael Kay</act2:name>
  </act2:actor>
 </act2:actors>
 <act2:institutions>
  <act2:institution id="I1">
   <act2:name>Saxonica</act2:name>
  </act2:institution>
 </act2:institutions>
 <act2:functions>
  <act2:function actor="A1" institution="I1">
   <act2:name>CEO</act2:name>
  </act2:function>
 </act2:functions>
</act2:database>
```

**Figure 5 - Document using a relationship**

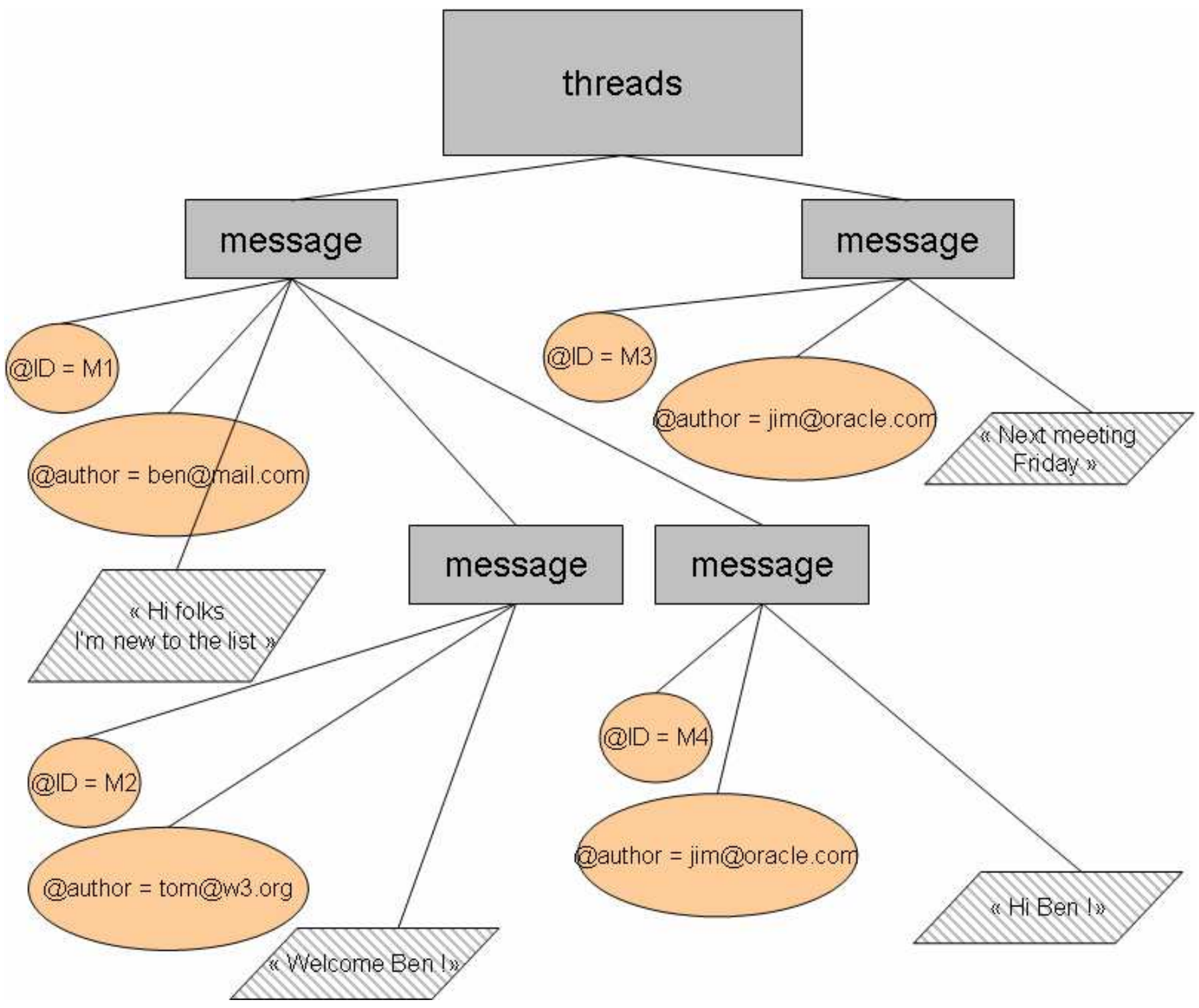

**Figure 6 - Recursive structure of messages**

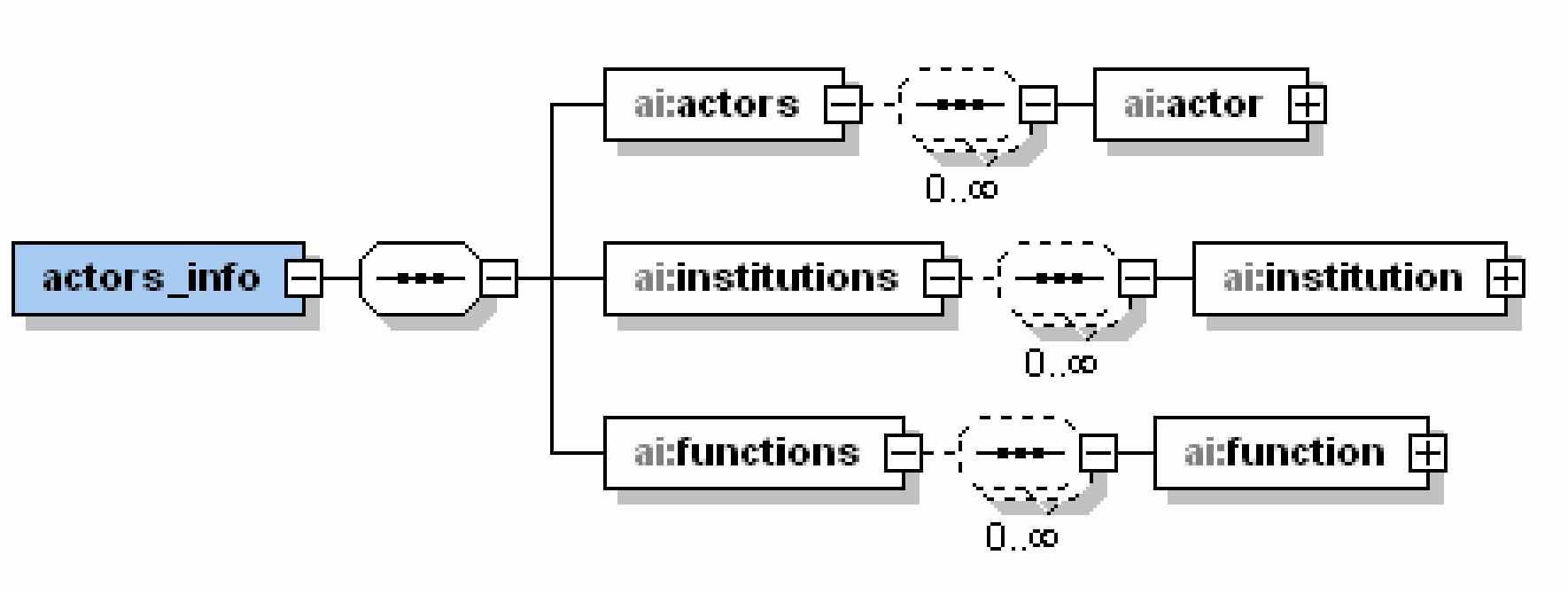

**Figure 7 - Global overview of the *actors_info.xsd* schema**

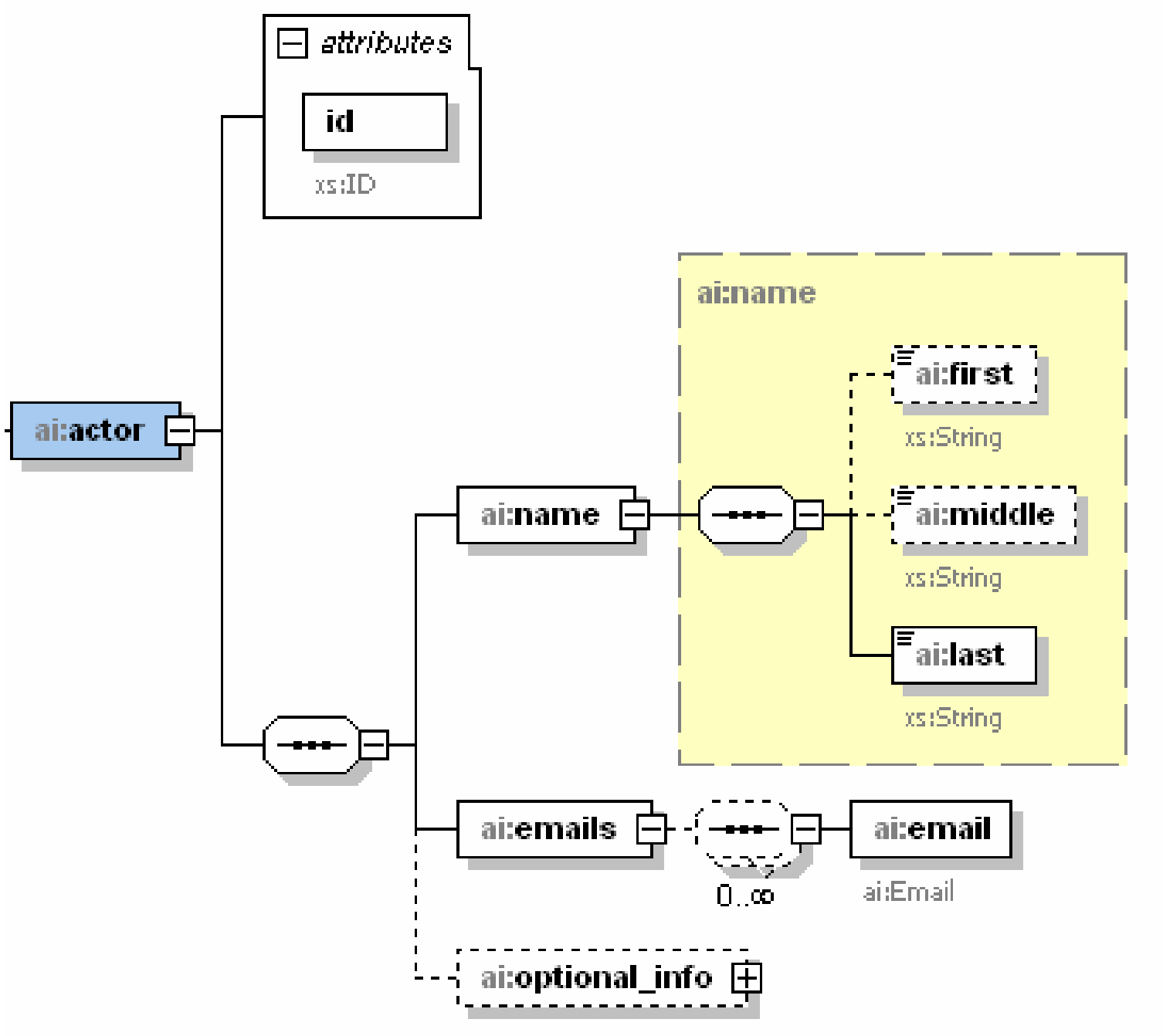

**Figure 8 - The *ai:actor* entity**

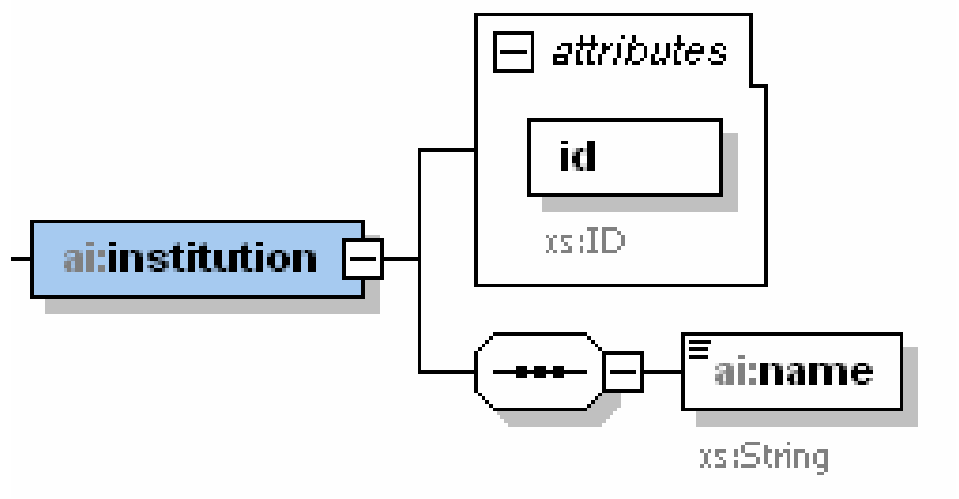

**Figure 9 - The *ai:institution* entity**

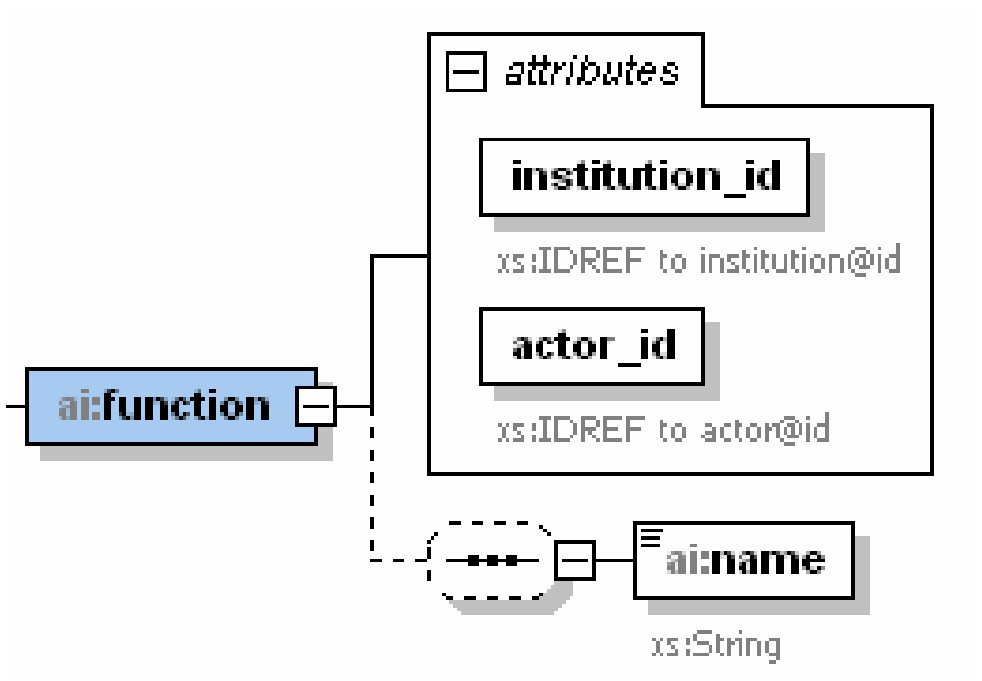

**Figure 10 - The *ai:function* relationship**

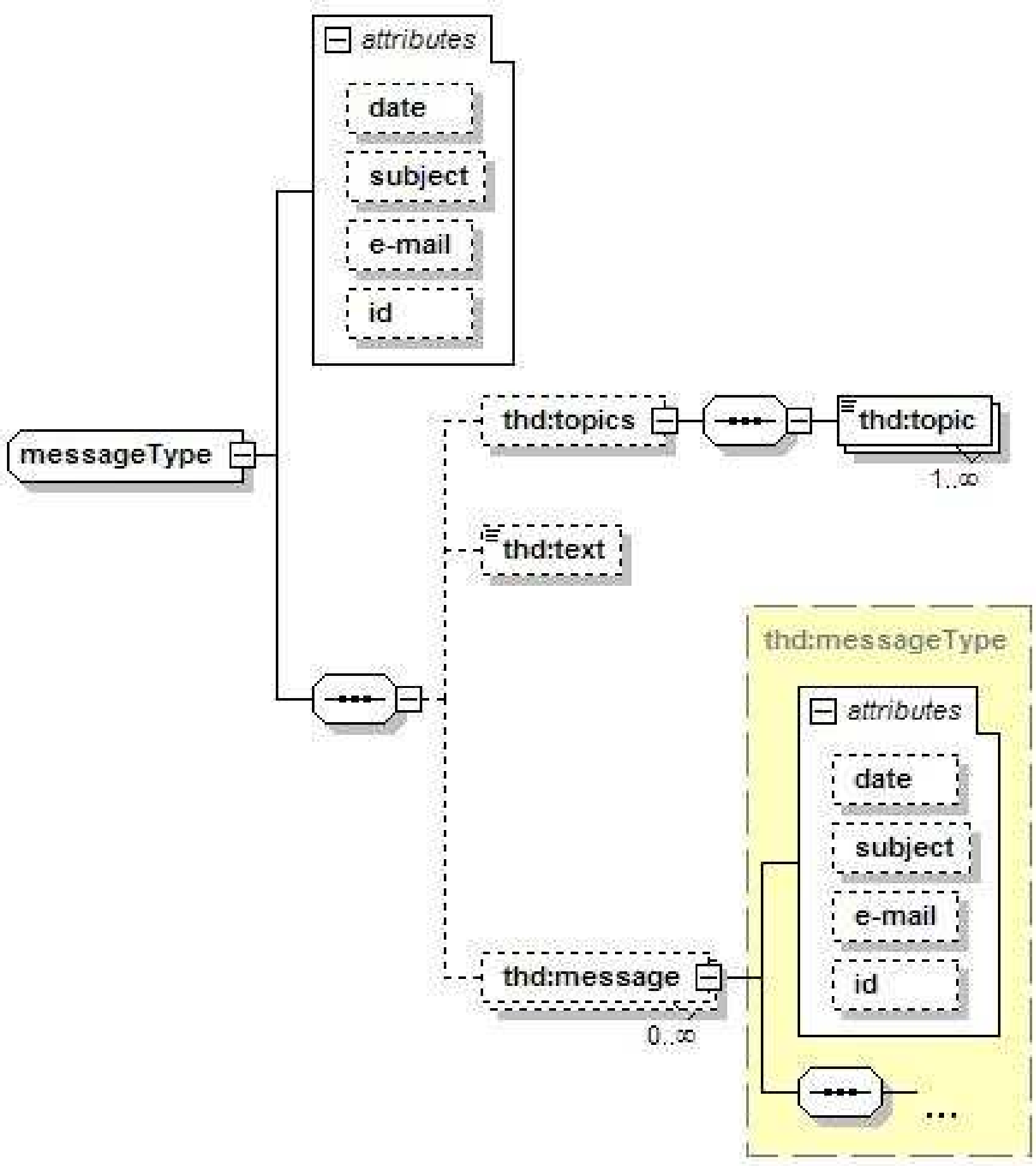

**Figure 11 - *Threads*.xsd : the recursive tree structure used to store the messages**

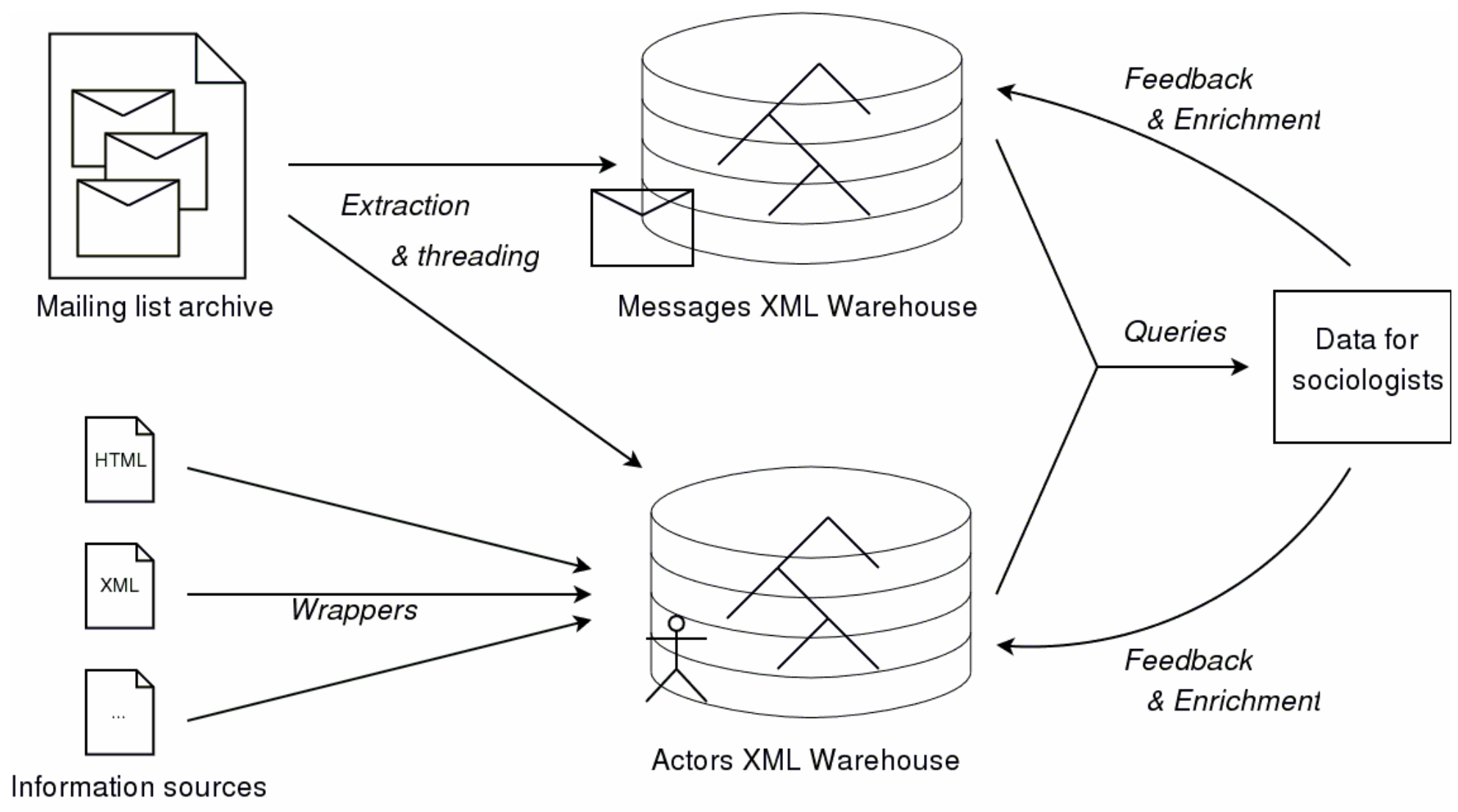

**Figure 12 - Data extraction process**

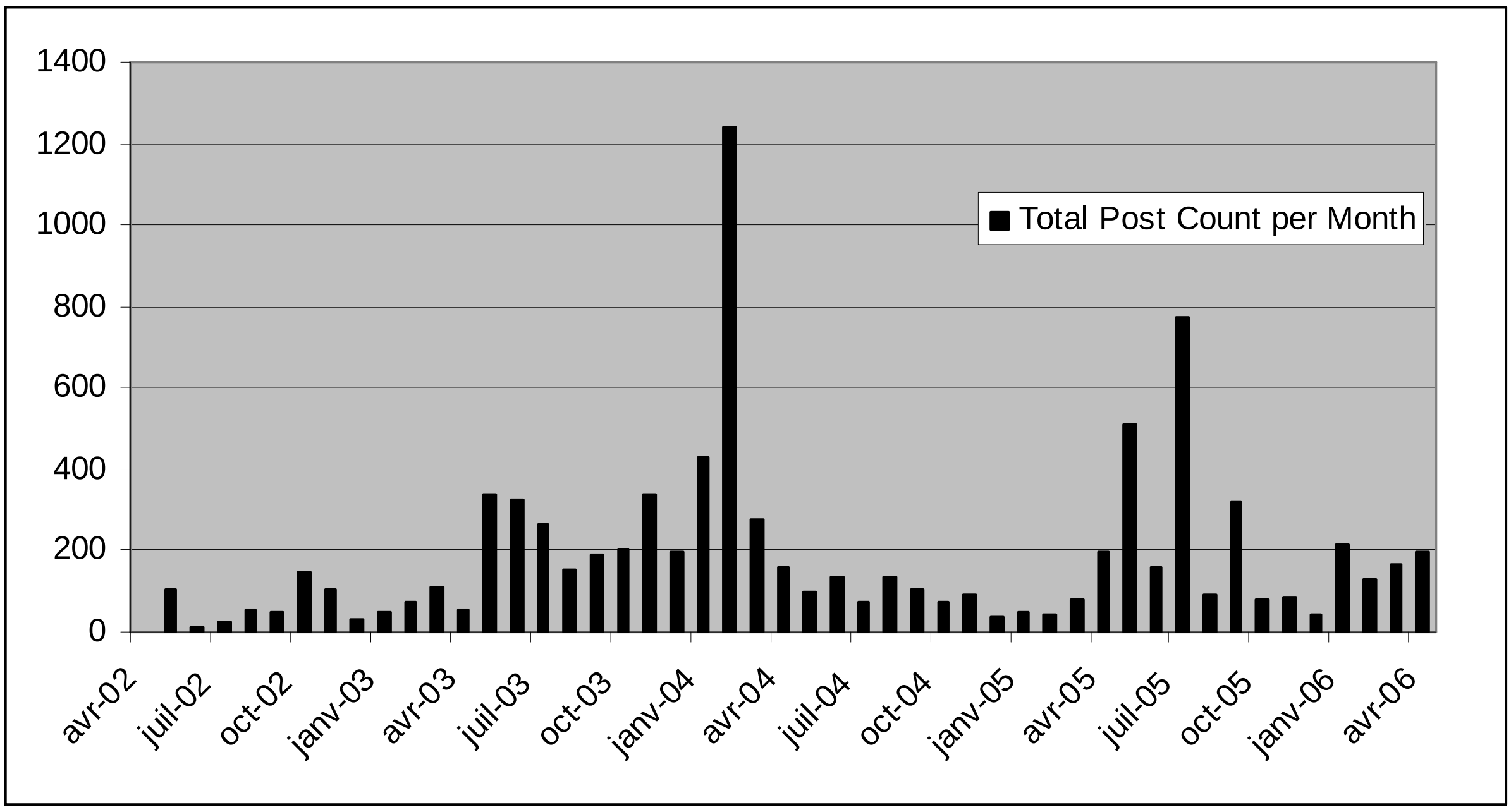

**Figure 13- Total number of emails poster per month from 2002 to 2006**

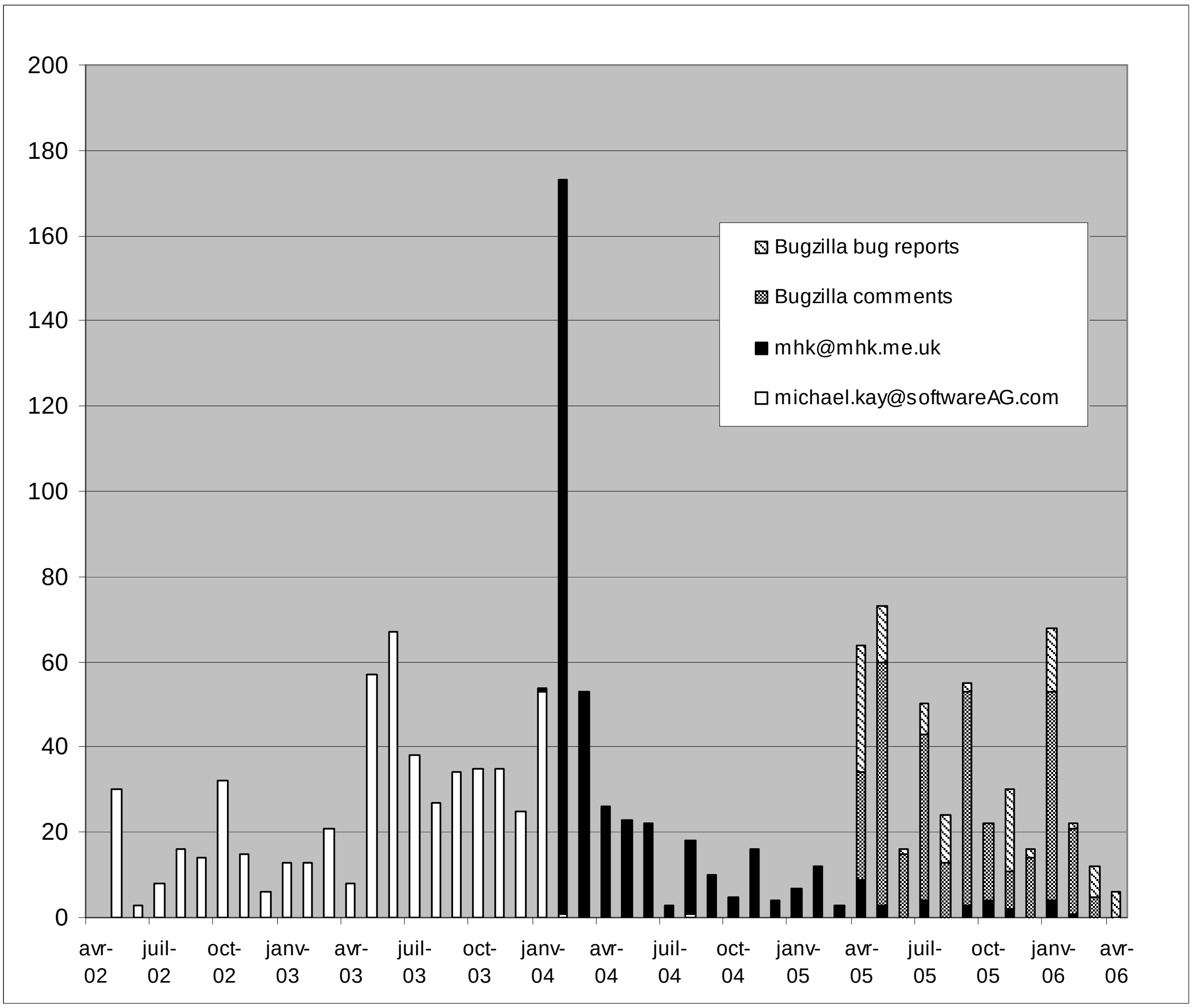

**Figure 14- Kay's postings per month**

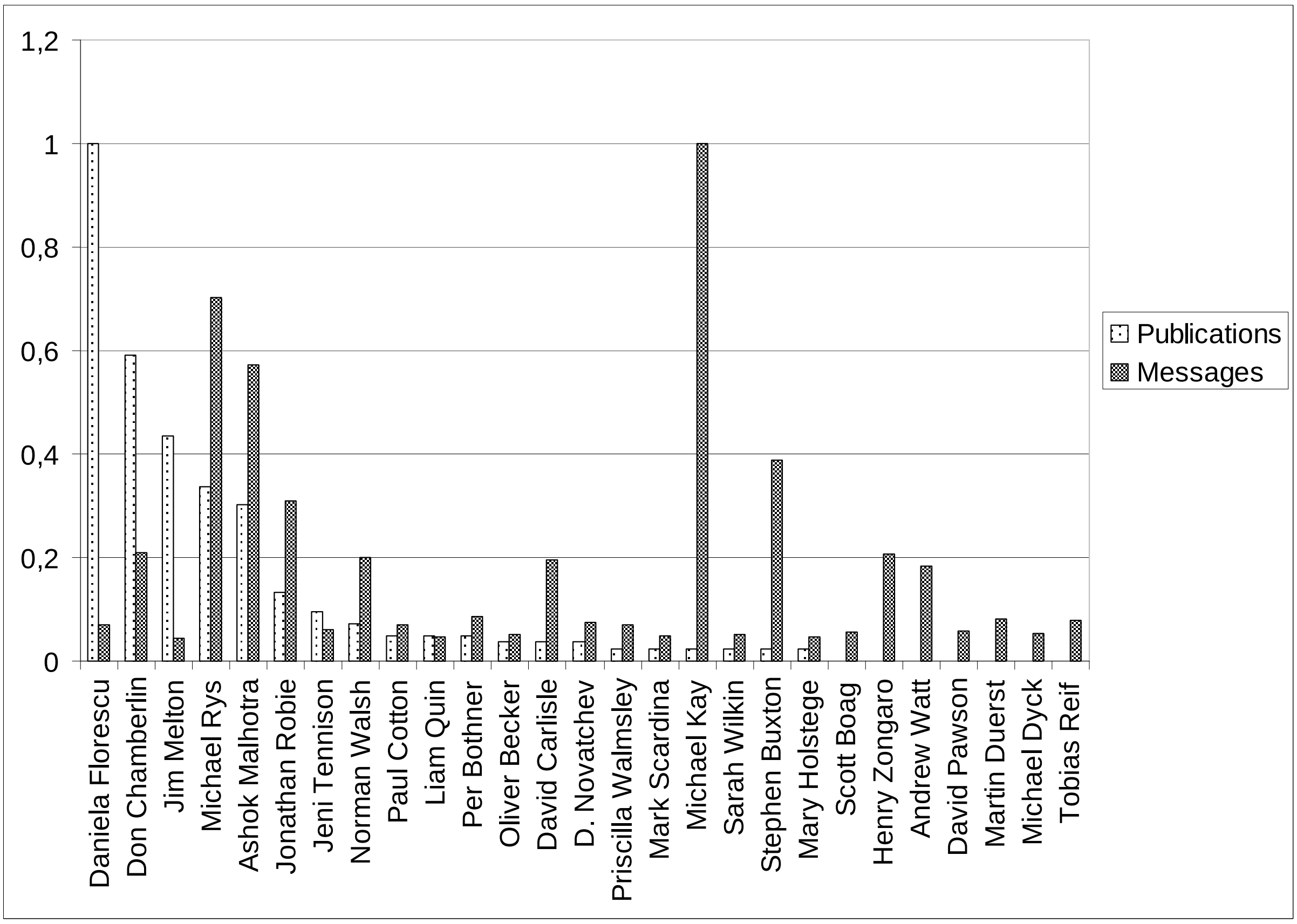

**Figure 15- List activism and scientific publication for top posters**

| | |
|---|---|
| microsoft.com | 1547 |
| ibm.com | 978 |
| softwareag.com | 681 |
| w3.org | 623 |
| oracle.com | 564 |
| cogsci.ed.ac.uk | 555 |
| acm.org | 485 |
| mhk.me.uk (saxonica) | 425 |
| nag.co.uk | 318 |
| yahoo.com | 288 |
| aol.com | 259 |
| datadirect.com | 212 |
| sun.com | 206 |
| arbortext.com | 203 |
| metalab.unc.edu | 196 |
| CraneSoftwrights.com | 180 |
| hotmail.com | 168 |
| kp.org | 165 |
| jclark.com | 141 |
| bea.com | 125 |

**Table 1- Top 20 posting institutions**